\documentclass[global,twocolumn,showpcs,amsmath,amssymb]{svjour}

\usepackage{graphics}% Include figure files
\usepackage{dcolumn}% Align table columns on decimal point
\usepackage{bm}% bold math
\usepackage{amsmath}
\usepackage{amssymb}
\usepackage{float}
\usepackage{bm}
\usepackage{array}
\usepackage{color}

\journalname{myjournal}
\begin{document}

\title{Compact, robust, and spectrally pure diode-laser system with a filtered output and a tunable copy for absolute referencing}

\author{E. Kirilov\inst{1} \and M.~J. Mark\inst{1} \and M. Segl\inst{1} \and H.-C. N\"{a}gerl
\inst{1}
\thanks{H.-C. N\"{a}gerl
 e-mail: christoph.naegerl@uibk.ac.at}%
}

\institute{Institut f\"{u}r Experimentalphysik und Zentrum f\"ur Quantenphysik, Universit\"at Innsbruck, Technikerstrasse 25/4, A-6020 Innsbruck, Austria}

\date{Received: date / Revised version: date}
\maketitle

\begin{abstract}
We report on a design of a compact laser system composed of an extended cavity diode laser with high passive stability and a pre-filter Fabri-Perot cavity. The laser is frequency stabilized relative to the cavity using a serrodyne technique with a correction bandwidth of $\geq 6$ MHz and a dynamic range of $\geq 700$ MHz. The free running laser system has a power spectral density (PSD) $\leq 100$ Hz$^{2}$/Hz centered mainly in the acoustic frequency range. A highly tunable, $0.5-1.3$ GHz copy of the spectrally pure output beam is provided, which can be used for further stabilization of the laser system to an ultra-stable reference. We demonstrate a simple one-channel lock to such a reference that brings down the PSD to the sub-Hz level. The tuning, frequency stabilization and sideband imprinting is achieved by a minimum number of key elements comprising a fibered EOM (electro-optic modulator), AOM (acousto-optic modulator) and a NLTL (non-linear transmission line). The system is easy to operate, scalable, and highly applicable to atomic/molecular experiments demanding high spectral purity, long-term stability, and robustness.
\end{abstract}
%\newpage
\maketitle

\section{Introduction}
\label{intro}
Stable and spectrally pure lasers have widespread applications in optical frequency standards \cite{Diddams2001}, measurement of fundamental constants \cite{Wineland2010,Marion2003,Eisele_gravity}, gravitational-wave detection \cite{Abbott2009}, quantum information \cite{Schindler2013}, high-resolution spectroscopy \cite{Kolachevsky2011} as well as stimulated adiabatic transfer of population between atomic/molecular levels \cite{Danzl2008,Ni2008,Mark2009,Danzl2010a,Aikawa2011a,Takekoshi2014}. It has already been demonstrated that the laser frequency can be locked to better than $0.1\,$Hz precision relative to a reference Fabri-Perot (FP) cavity. The ultimate linewidth achieved is then limited by the cavity's mechanical vibrational and thermal noise. Recently a linewidth of $<40\,$mHz  was demonstrated by stabilizing a erbium-doped fiber laser to a silicon single-crystal Fabri-Perot cavity held at cryogenic temperatures \cite{Kessler2012}.

Diode lasers are generally compact and cost effective, cover wide spectral ranges, can be tuned to the desired wavelength, and can be optically narrowed. For an extended cavity diode laser (ECDL) with the usual grating feedback the linewidth easily narrows to the $200\,$kHz range \cite{Riehle2004}. To further reduce the linewidth additional optical feedback from an external cavity \cite{Breant1989} can be employed, combined with a low frequency electronic lock to a reference cavity (RC) \cite{Kirilov2009}. The optical lock is difficult to maintain since one has to stabilize the ``half-cavity" between the ECDL and the optical feedback FP. Different variations to stabilize and increase the dynamic range of the optical lock include balanced polarization-sensitive methods or dithering of the half-cavity path \cite{Doringshoff2007,Labaziewicz2007}. For absolute frequency reference and additional noise reduction the RC spacer is fabricated out of ULE (Corning), Zerodur, or single-crystal silicon substrates, and the laser is stabilized at the zero crossing of the spacer's thermal expansion coefficient. The error signal from referencing the laser to the RC is generated using the Pound-Drever-Hall (PDH) technique \cite{Black2001}. In almost all scenarios for tuning the laser frequency in the range of the free spectral range (FSR) of the RC a configurations of at least 2 double-pass AOMs has to be employed.

As an alternative to external cavity optical feedback one can lengthen the cavity length $L$ of the ECDL \cite{Kolachevsky2011}, since the linewidth $\Delta\nu$ depends inversely quadratically on the length as $\Delta\nu=\Delta\nu_\text{LD}/(1+L/L_\text{LD})^{2}$, where $L_\text{LD} $ is the optical length of the laser diode (LD). In addition, the laser is typically locked to a RC with a high bandwidth (BW) electronic lock. The mode-hop free tunability is then compromised relative to a short cavity setup. Typically in most designs the slow integrated correction signal is applied to the piezo element (PZT) regulating the ECDL cavity length and the fast signal is applied to the injection current of the laser diode. The fast channel BW is typically limited to $1-2\,$MHz due to competing thermal and electronic charge carrier effects with opposite signs of $d\lambda/dI$, where $\lambda$ is the laser wavelength and $I$ is the diode injection current. This is specific to the respective laser diode and requires that one individually adapts the design of the proportional-integral-derivative (PID) servo. In some of the above designs the useful output sent to the experiment is taken in transmission through a pre-stabilizing filter cavity (FC) (which could be the same one providing the optical feedback \cite{Labaziewicz2007}) to erase the remaining noise pedestal due to stochastic laser phase modulation, which typically caries about $10\%$ of the laser power and whose width $\sim 1\,$MHz and shape depend crucially on the PID parameters \cite{Matveev2008,yatsenko2014}.

Here we present a diode-laser system that is extremely robust to acoustic perturbations and that has superior long-term stability with high modularity suitable for various demanding applications. The linewidth narrowing is accomplished by a method independent of the specific laser diode. The continuous tunability of the laser without feed-forward \cite{Petridis2001} is $\sim 3\,$GHz. We employ a low-acoustically susceptible mechanical design of an ECDL in Littrow configuration in combination with a high BW and dynamic range fibered EOM-NLTL \cite{Kohlhaas2012,Afshari2005,Johnson2010a}, a single doubly-diffracting AOM, and a FC, each contributing both to the tuning and stabilization of the laser. The setup can be used in many different, more familiar configurations that are easily interchangeable from the one described in this work, where the roles of the components are disentangled, similar but still superior to standard laser systems, with the cost of adding additional infrastructure. The passive stability of the ECDL combined with the features of the above elements provides an economic and elegant solution to a high performance laser system and a good alternative to existing designs.

\section{Basic description of the laser system. Advantages of the design}
\label{sec:1}
\begin{figure}	
\includegraphics{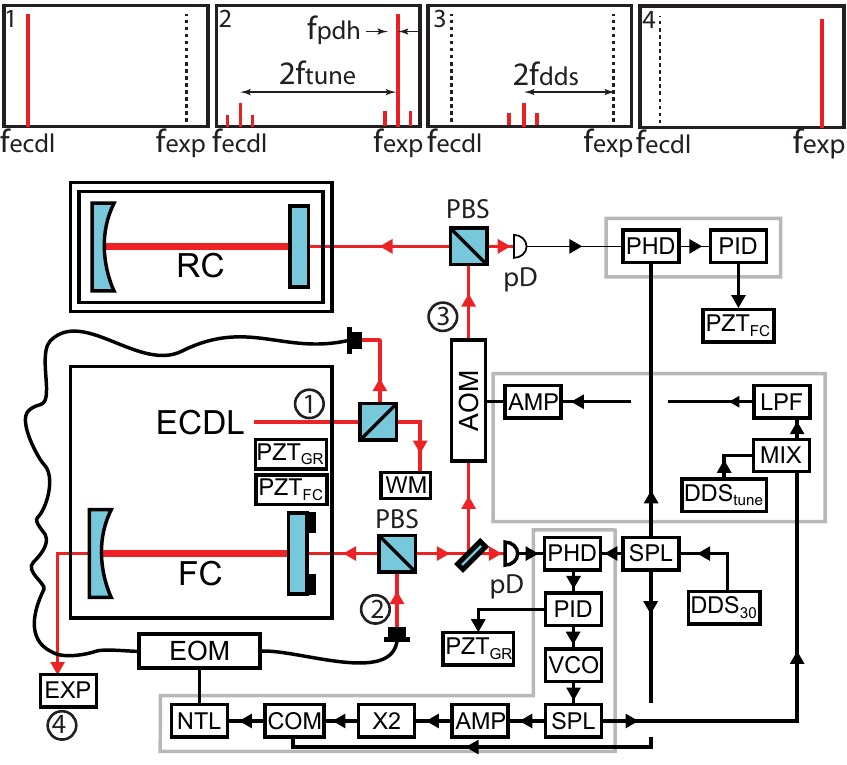}
\caption{Overall setup. Panels 1 through 4 indicate the frequency components present at different locations in the setup as indicated. Notation that is not defined in the text: PHD: phase detector, SPL: radio-frequency (rf) power splitter, COM: rf combiner, AMP: rf low power ($\textless 25$) dBm amplifier, X2: rf doubler, MIX: rf mixer, LPF: low pass filter, PBS: polarizing beamsplitter, PD: photodiode.}\label{full_diagram_0}
\end{figure}

The rough scheme is presented on Fig.\ref{full_diagram_0}. The output of the ECDL is passed through a fibered EOM driven by the superposition of a direct digital synthesizer's (DDS) sinusoidal output at fixed frequency $f_\text{pdh}$ and a sawtooth signal generated by a NLTL powered by a frequency-doubled broadband high-bandwidth voltage-controlled oscillator (VCO) at $2f_\text{tune}\gg f_\text{pdh}$ (the factor of 2 results from a doubling stage after the VCO; see below). Most of the light entering the EOM at frequency $f_\text{ecdl}$ is diffracted into the upper (or lower, but not both; a property of the serrodyne modulation technique \cite{Johnson2010a}, details in sec.\ref{sec:2}) sideband at $2f_\text{tune}$. The spectral distribution after the EOM is composed of mostly $f_\text{ecdl}+2f_\text{tune}$ and a small leftover at $f_\text{ecdl}$ with both frequencies accompanied by their small sidebands at $\pm f_\text{pdh}$ (panel 1 of Fig.\ref{full_diagram_0}). Further the light is directed towards a 2-port tunable FC (tuning is possible by a PZT) that is resonant with $f_\text{ecdl}+2f_\text{tune}$. The laser is locked to the FC using a PDH lock at $f_\text{pdh}$ by slow/fast lock sent to the ECDL-PZT/EOM-NLTL respectively. In transmission the filtered light goes to the experiment with frequency $f_\text{exp}=f_\text{ecdl}+2f_\text{tune}$ (panel 4 of Fig.\ref{full_diagram_0}). In reflection the light consists now of mostly $f_\text{ecdl}$ and a leftover at $f_\text{ecdl}+2f_\text{tune}$ (the last portion results from the non-perfect mode matching and mirror losses). A small portion of it is necessary for the PDH lock (using the lines at $f_\text{ecdl}+2f_\text{tune}\pm f_\text{pdh}$), but the rest is passed through a double-pass up-shifting AOM. The AOM is driven by the frequency difference $f_\text{tune}-f_\text{dds}$, where the first portion comes from the same VCO driven by the fast port of the PID used to lock to the FC, but before the doubling stage, (Fig.\ref{full_diagram_0}; denoted X2) and the second from a DDS (subtracted using a rf mixer; denoted MIX). The useful light after the double-pass AOM has the frequency $f_\text{ecdl}+2f_\text{tune}-2f_\text{dds}\pm f_\text{pdh}$ (panel 3 of Fig.\ref{full_diagram_0}). The result is a filtered, ultra-tightly locked main beam coming in transmission of the FC with noise dominated only by the cavity's acoustics, and another beam (the one after the AOM) that is shifted $2f_\text{dds}$ away, which, in the acoustic frequency range, perfectly resembles the primary one. The latter one can be used to lock the whole system to an ultimate RC by a simple one channel PID with mediocre bandwidth, addressing the PZT of the FC.\\
The system's advantages apart from the above mentioned ones over the standard realizations of such systems include: 1) extremely high BW and dynamic range of the fibered EOM-NLTL system superior to intra-cavity EOMs (which have smaller dynamic range) or injection current locks (which have smaller BW and dynamic range and add amplitude noise), 2) comparatively low optical losses (mainly due to the NLTL), 3) convenient long-range tuning without the need to optimize AOMs when switching to another atomic or molecular line as one simply has to change the DC value $f_\text{tune}$ and accordingly $f_\text{dds}$ to be in the convenient range of the AOM (the fast lock to the EOM is ac-coupled starting from $>10\,$Hz), 4) decreased complexity by a) achieving tuning, sideband modulation, and locking (to both FC and RC) by just one EOM and one AOM, b) powering all the components by $\textless 25\,$dBm rf amplifiers (still in the power range where the costs are minimal), and c) using only one actuator, namely the FC PZT, to lock to the RC in view of its high BW (a small mirror is used and a carefully chosen PZT), and 5) filtered output by the FC, eliminating stochastic laser noise and unwanted sidebands.

\begin{figure}	
\resizebox{0.5\textwidth}{!}{
\includegraphics{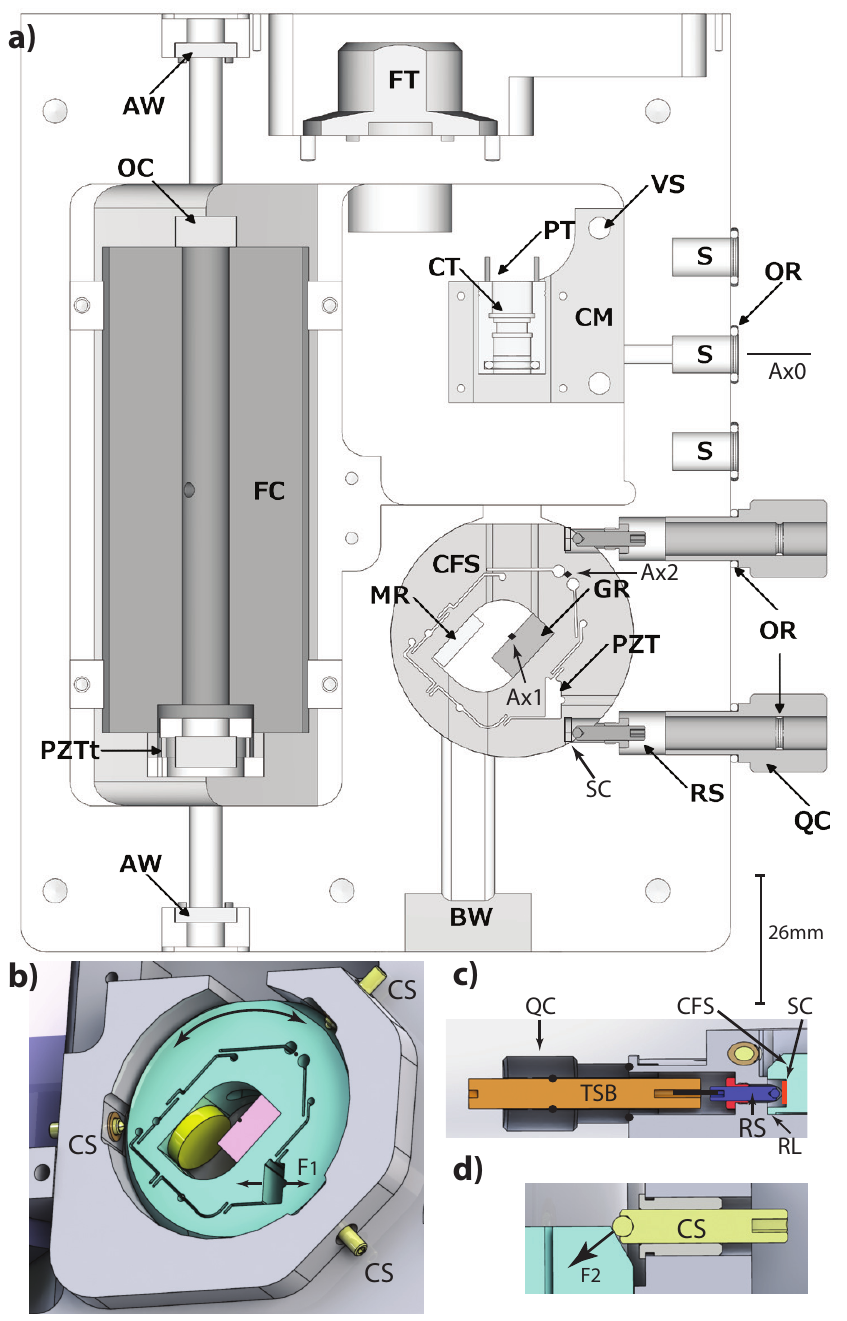}
}
\caption{a) ECDL design. Top view cross-section at about mid plane. Notation: CFS: cylinder-flexure structure, CM: collimator mount goniometer, AW: anti-reflection coated windows (serving also as vacuum seal with viton O-rings), PZT: grating piezo, PZTt: two tubular concentric PZTs, FC: filter cavity, OC: output coupler, BW: brewster window, PT: Peltier elements, CT: collimator tube for laser diode, VS: precision screws allowing for vertical alignment of the diode (by pressing against the floor of the laser bed), RS: precision screws for rough wavelength tuning, CS: centering screws, SC: sapphire crystal, QC: vacuum quick connectors, OR: viton O-rings for vacuum seal, TSB: transmission-sealing bar, GR: holographic grating, MR: mirror for steering the beam, S: clearance holes for holding the CM to the laser body, FT: electrical feedthrough. Ax0: axis of rotation of CM, Ax1: axis of rotation of CFS driven by RS for rough wavelength selection, Ax2: axis with zero velocity during flexing driven by PZT, F1: force exerted by PZT, F2: force exerted on the CFS by each CS. b) 3D view of the CFS with the three CS providing smooth rotation. c) Zoom in of the vertical cross section of the rough CFS rotation driving mechanism. RL: ring shaped lip of the CFS. d) Zoom in of the upper round cleaved edge of the CFS experiencing a force F2 by the ball tip of the CS. The scale bar refers only to a). }
\label{full_anatomy1}
\end{figure}
	
\section{ECDL engineering and performance}
\label{sec:2}
We first discuss the laser housing and mount in detail. It is based on a Littrow-type ECDL with a variable cavity length in the range of $3-10\,$cm. The passive stability of the ECDL (operating with a laser diode EYP-RWE-1060, Eagleyard) and acoustic immunity are improved in comparison to standard designs by a few simple measures (Fig.\ref{full_anatomy1}). First, a commercial diode collimation tube (LT230P-B, Thorlabs) is partially flattened on two parallel planes and sandwiched between two Peltier elements (TE-17-1.0-2.0, TE tech.) to control the diode's temperature using a $10$ k$\Omega$ thermistor (Vishay BCComponents). The overall square-shaped mount (CM) serves as a heat sink and as a vertical goniometer. Rotation in the vertical plane (around axis Ax0) is necessary to optimize the optical feedback. It is initiated by vertical precision screws (VS) pressing against the floor of the main body. Once aligned, the vertical rotation degree of freedom is frozen by tightening the CM to the sidewall of the laser body \cite{Cook2012} as it is not used in a daily operation. Its contribution therefore to the acoustic noise of the laser is minimized. The grating (holographic, number of lines/mm chosen such that $\theta_\text{littrow}\sim 45^{\circ}$; 33025FL01-239H Richardson gratings) is positioned in a cylinder-flexure structure (CFS) that is allowed to rotate only in the horizontal plane for tuning the wavelength (around Ax1) \cite{Toptica}. Rough tuning is achieved by a rotation induced by two fine threaded screws (RS) (F3SS10 and N250L3P, Thorlabs) inducing rotation in opposite directions by pushing along the tangent of the CFS (see Fig.\ref{full_anatomy1} a) and c) for horizontal and vertical views respectively). The screws have ball tips and press against the CFS through two sapphire crystals (SC, 43-627, Edmund). Smooth rotation is accomplished by three ball-tip precise screws (CS) that press the upper edge of the cylinder, which is wedged at $30^{\circ}$ (Fig.\ref{full_anatomy1} d)). This feature also assures that the screws apply a normal downward force component to the cylinder (F2), therefore guaranteeing rotation only in the horizontal plane. The CFS contacts the main body only through the ball tips of the three CS and a thin 1 mm wide ring-shaped lip left on the outer edge of its bottom surface (denoted RL and visible in Fig.\ref{full_anatomy1} c)). After choosing the rough wavelength the horizontal rotation is also frozen (by applying a force on the CFS by both RS). Subsequent fine tuning is left to a PZT stack (PCh 150/7x7/2, Piezomechanik) driving only the flexure (force F1 on Fig.\ref{full_anatomy1} b) ), designed to have a mechanical frequency at $\sim 15\,$kHz. The center of the grating moves $\sim 1 \, \mu$m for $100\,$N force applied by the PZT. During flexing the grating experiences a rotation around Ax2. The CFS is made out of 316 SS. Modeling of the above properties is achieved using a standard CAD program with an elastic properties simulation capability. In practice the resonant frequency is determined by exciting the PZT and sweeping a variable sound generator while observing the noise spectral density of the laser when locked to the FC. The whole body of the laser is made out of aluminum and is vacuum sealed, allowing it to be evacuated if desired. It is thermally stabilized by 4 Peltier elements (TE-127-1.0-1.3, Tetech.) wired in series. The resistive bridges for both the diode and laser body are attached on the back of the laser for thermal stability and share the layout with the LD filtering circuit. Rough tuning is possible by engaging the two finely threaded screws RS with adapted quick-seal connects QC (B-025-K, Lesker), which allows for rotation without breaking vacuum (applying Apiezon vacuum grease on the sealing o-rings experiencing the rotation driven by transmission-seal bar (TSB) from outside). The chamber is evacuated through a small inlet (same as V1021-1, DLH Industries) machined on a KF25 flange, which is economic and does not take much space compared to standard vacuum valves.

The performance of the bare, free-running ECDL without engaging the tight locks to the FC and the RC is characterized by analyzing the frequency noise linear spectral density (LSD) of the error signal as shown in Fig.\ref{noise_comparison} when the laser is locked to the FC using only the slow ($\sim 10\,$Hz) servo branch applied to the PZT behind the grating of the ECDL (the feedback via the EOM is not applied). There are no visible mechanical resonances. The ``fast" linewidth of the laser is $\leq 20\,$kHz mostly defined by the length of the ECDL cavity. At Fourier frequencies $\leq 100\,$kHz the spectrum increases from the quantum white noise level due to additional technical noise sources. The overall rms linewidth is $\Delta\nu^\text{ecdl}_\text{rms}=\int_{0}^{\infty}S_{\Delta\nu}(f)df=59 \,$kHz. The corresponding Allan deviation calculated based on the PSD $\sigma^{2}(2,\tau)(\text{Hz}^2)=2\int_{0}^{\infty}S_{\Delta\nu}(f)\sin{(f\tau)}^{4}/(\pi f\tau)^{2}df$ is in the $\leq 25 \,$kHz range for times in the $10 \, \mu$s - $10$ ms interval.

\section{Full laser system description and performance}
\label{sec:3}
\begin{figure*}	
\resizebox{1\textwidth}{!}{
\includegraphics{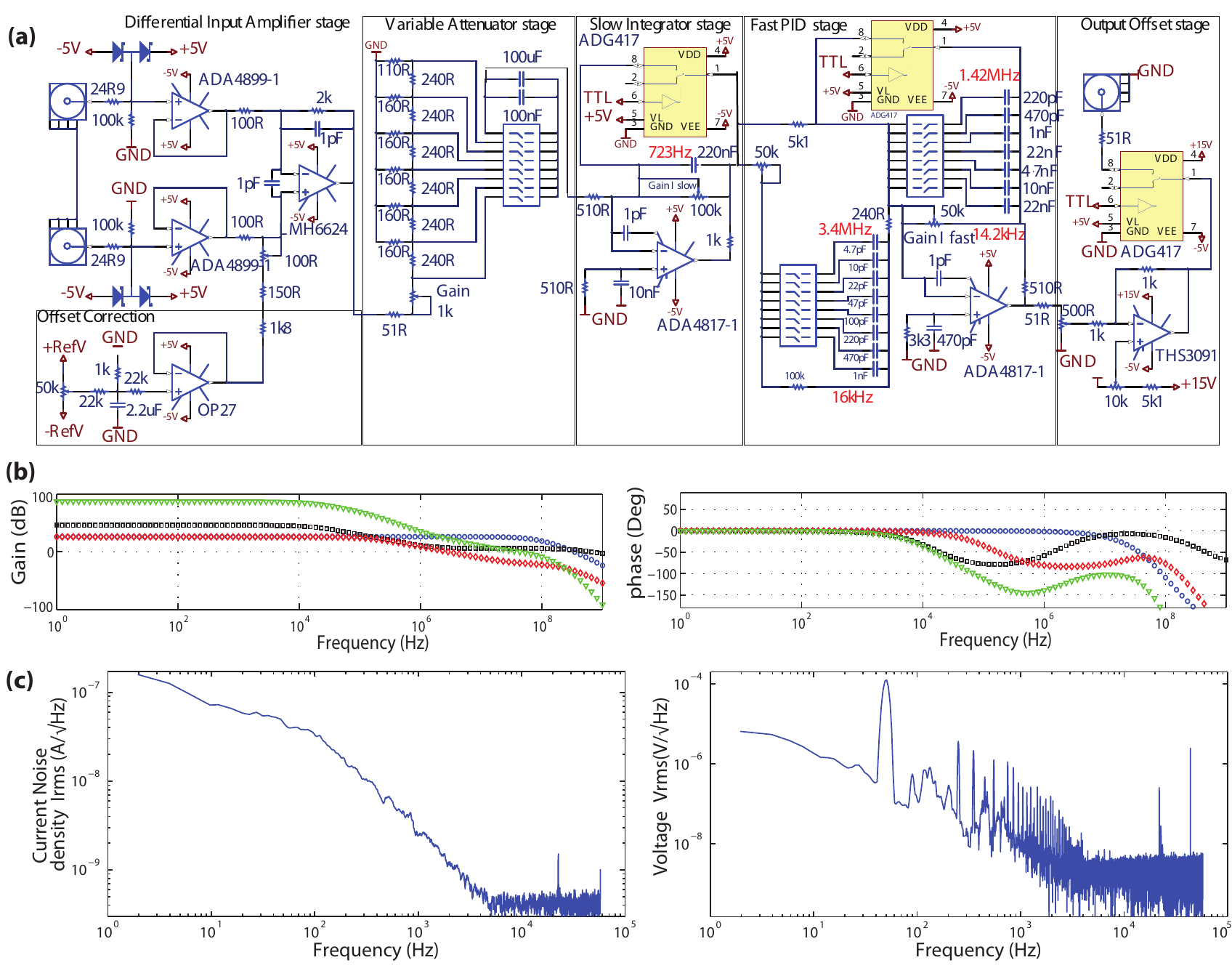}
}
\caption{a) PID circuit for the fast correction fed to the EOM. Frequencies in red are examples for the components values used in this specific circuit and may be modified depending on the actuators BW for the specific laser system. b) Simulated Bode plots of the PID circuit with typical parameters of operation; slow integrator (black squares), fast integrator (red diamonds), op. amps. (blue circles), full PID (green triangles), c) LSD of the LD current source and the PZT driver.}\label{full_diagram_electronics}
\end{figure*}

We now use the fibered EOM (EOspace, PM-0K5-10-PFA-PFA-980, 10 GHz BW) as a fast actuator. Almost $50\%$ of the light is coupled into the EOM, additionally spatially filtering the beam and thus delivering a clean Gaussian mode. We note that one typically filters the laser mode with a polarization maintaining (PM) fiber; the presence of the fibered EOM therefore eliminates this step. The light is then directed to the plano-convex FC (finesse $\sim 9400$, FSR $= 1.481$ GHz, spacer made of Zerodur square block 30x30x100 mm, Helma optics) with one coupler (both couplers from Layertec, low loss, 12.5 mm diam.; flat and $500$ mm radius) mounted on two concentric tubular PZT's (Meggit, F3270154, F3270055) compensating their thermal drifts (Fig.\ref{full_anatomy1}; PZTt). The FC is placed in a V-groove and held in place by 8 viton O-rings ($3\,$ mm OD) positioned closely under and above the diagonal-mid plane (top clamps visible on Fig.\ref{full_anatomy1}). Almost $80\%$ of the light is coupled into the FC on resonance. The EOM is driven by a NLTL 7013 (Tactron Inc., produced by Picosecond Pulse Inc., operating range $400-1600 \,$ MHz, higher than the specifications), which converts the signal from the VCO (VCO ROS-625-219+, Mini-Circuits) into a sawtooth-shaped signal \cite{Johnson2010a}. When Fourier transformed the signal of the NLTL driven at frequency $f$ is composed of a comb of frequencies with power scaling roughly as $\sim 1/n$, where $n$ is the Fourier index. Once the amplitude of the sawtooth signal becomes equal to the $V_{\pi}$ voltage of the EOM (i.e. 3-4 V), the modulated phase has an effectively linear time dependence and therefore most of the light power ($\sim 70\%$) is diffracted either into the upper or into the lower sideband spaced at $2f_{tune}$ (500-1300 MHz) away from the carrier. This sideband can be tuned with $3$ dB BW of $\sim 25$ MHz limited by the VCO BW and with a dynamic range of $\sim 800$ MHz. This allows for an extremely fast and easy to operate lock with a closed loop BW that can be pushed to $8$ MHz, limited primarily by phase delays in the PID (see Fig.\ref{full_diagram_electronics} a) and c) for a schematic) and cable length.

\begin{figure}
\includegraphics{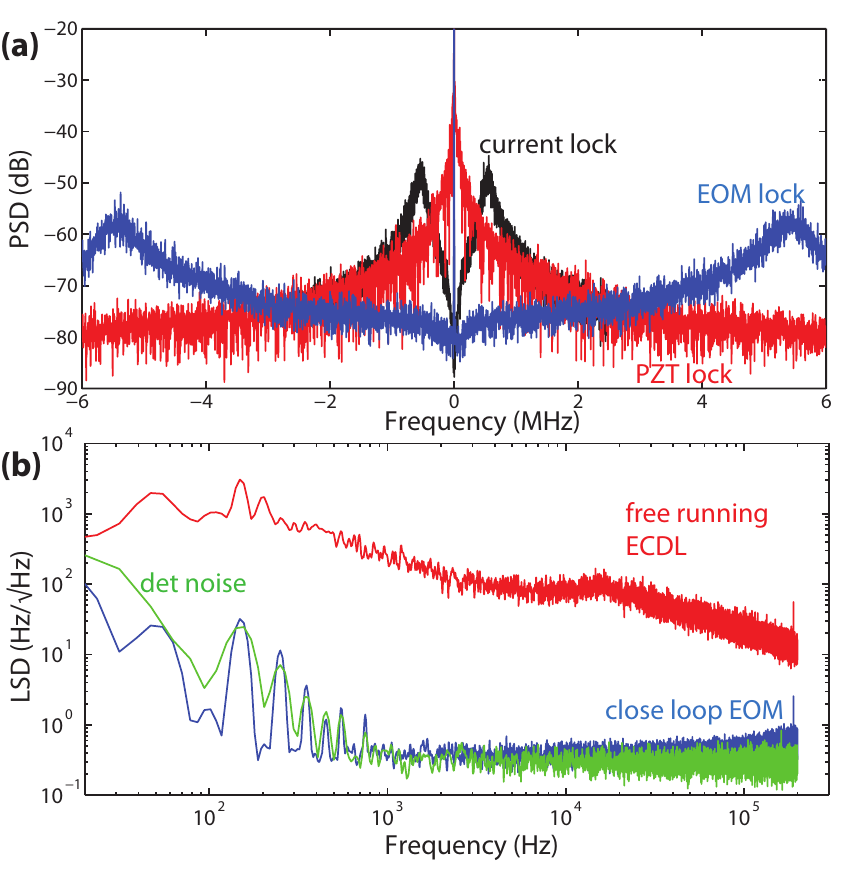}
\caption{Characterization of the laser noise performance. a) Frequency noise power spectral densities (PSD) of the PDH error signal as recorded by a spectrum analyzer for the laser stabilized to the FC for 1) just a PZT slow lock not to influence the PSD at frequencies $>10$ Hz, 2) applying a fast lock to the injection current, and 3) adding the fast lock to the fibered EOM (signal taken before the phase detector and then translated to zero frequency). b) LSD of the PDH signal as measured by an audio analyzer for the cases as indicated.}\label{noise_comparison}
\end{figure}

The FC is positioned in the same body as the ECDL to minimize space and costs. The whole chamber is covered with silicon (Smooth-on Mold Max 15T) for damping acoustic air-born noise \cite{Cook2012}. The sidebands needed for the PDH lock are imprinted by a $30$-MHz wave provided by a fixed DDS (AD9858) right before the NLTL (power splitters resp. combiners are ADP-2-20-75, Mini-Circuits). These sidebands pass the NLTL unaltered. One could perform the mixing after the NLTL with a resistive combiner, but then a $6$ dB loss is added and the $V_{\pi}$ of the EOM can barely be reached. In reflection $10\%$ of the light is used for the PDH to the FC. The fast correction is applied to the voltage-controlled port of the VCO. The slow correction ($\sim 100$ Hz) goes to the grating piezo (PZT) of the ECDL. At this point the laser is locked to sub-Hz linewidth relative to the FC and the PSD is dominated solely by the FC's acoustic noise. Fig.\ref{noise_comparison} a) and b) compare the PSD/LSD of the error signal to the FC for the free running ECDL and in a locked configuration. For comparison we show the PSD when using the injection current of the LD as a fast actuator. We reach a unity gain frequency of $\sim 1$ MHz for the servo loop when locking via the current without lead compensation. Sending the fast correction to the fibered EOM effortlessly achieves unity gain frequencies that could be pushed to $\geq 8$ MHz. For low frequencies the noise cancellation is more than $50$ dB.

To achieve such high BW a careful design of the PID circuit is needed. In Fig.\ref{full_diagram_electronics} a) we provide the schematic and describe the most important features. The differential input amplifier stage consists of two unity gain buffers (ADA4899-1, $600$ MHz GBWP, $1\ \mathrm{nV}/\sqrt{\mathrm{Hz}}$ LSD) and a subtracting high BW amplifier (LMH6624, $1.5$ GHz GBWP, $0.92 \ \mathrm{nV}/\sqrt{\mathrm{Hz}}$ LSD) with a fixed voltage gain of $20$ dB. This configuration ensures a high input impedance optimal for signals coming from phase-detectors and a high usable signal BW while keeping the overall noise level low. Next a variable attenuator stage allows for a reduction of gain if necessary. Additionally one can choose between AC and DC coupling of the signal (corner frequency $\sim 10$ Hz). The first (slow) integrator stage has a fixed corner frequency of $10$ kHz. The maximum voltage gain for low frequencies reaches $46$ dB, the gain in the proportional region is fixed to $6$ dB. The amplifier that we use (ADA4817-1, $1$ GHz GBWP, $4 \ \mathrm{nV}/\sqrt{\mathrm{Hz}}$ LSD) has FET-input stages, ensuring negligible changes in the offset voltage in the whole integral gain region. The full PID stage, using the same amplifier as the slow integrator stage, uses DIP-switches to set the corner frequencies of the D- and I-part by changing the corresponding capacitances. The P-part has a fixed attenuation of $-20$ dB, ensuring that the D-part with a maximum of $6$ dB gain is within a frequency region where the bandwidth of the amplifier is no limitation. The available corner frequencies of the I-part with a maximum gain of $20$ dB are limited to low frequencies such that no overlap with the first (slow) integrator could lead to instabilities due to high phase shifts. A full-scale output offset can be added within the last stage (THS3091, $190$ MHz Bandwidth, $2 \ \mathrm{nV}/\sqrt{\mathrm{Hz}}$ LSD), which also serves as an output buffer able to provide up to $250$ mA current.

Overall, all electronic parts, especially the amplifiers, are chosen carefully to allow a maximum electronic bandwidth of several tens of MHz such that phase shifts due to the amplifiers of the PID circuit are negligible in the overall PID, thereby not limiting the feedback BW. Bode plots showing the performance of our circuitry are given in Fig. \ref{full_diagram_electronics}. To minimize electronic noise introduced by the PID, ultra-low-noise amplifiers and small resistor values to reduce thermal noise are used. The current and PZT driver circuits are also home made with noise figures of $\textless 32 \ \mu \mathrm{A}_{\mathrm{rms}}$ ($\textless 1.3 \ \mathrm{nA}/\sqrt{\mathrm{Hz}}$ @ $2$ kHz) for the current and $\textless 860 \ \mu \mathrm{V}_{\mathrm{rms}}$ ($\textless 5 \ \mathrm{nV}/\sqrt{\mathrm{Hz}}$ @ $2$ kHz) for the piezo in the $1$ Hz - $20$ kHz range (see Fig. \ref{full_diagram_electronics} c)).

\begin{figure}
\includegraphics{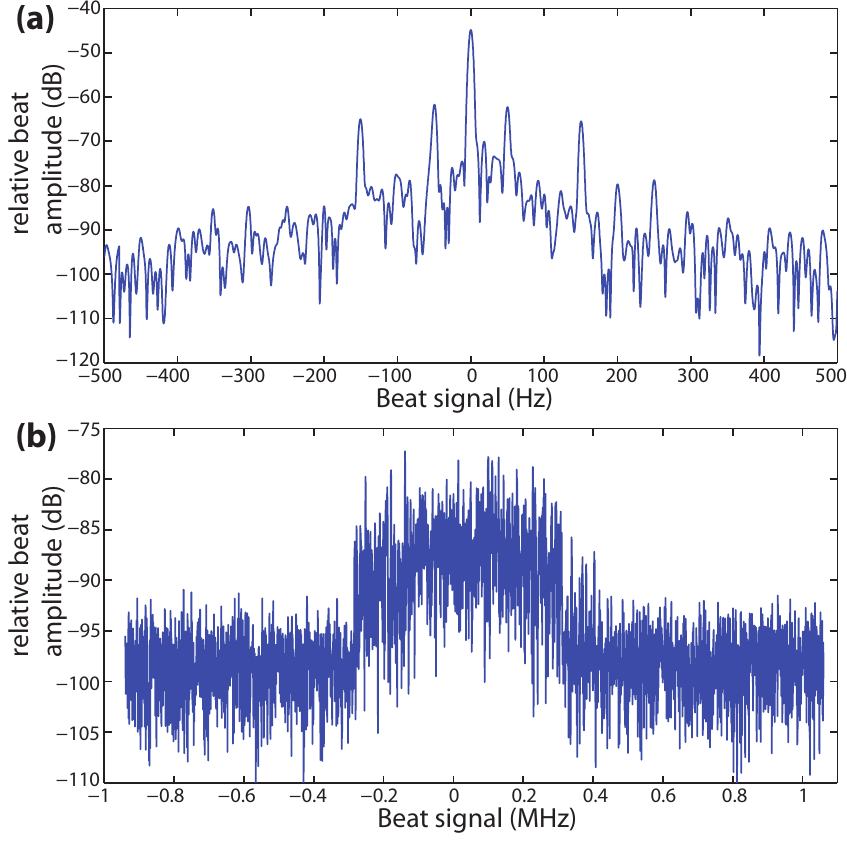}
\caption{Beat spectrum between a) the laser field in transmission of the FC and the laser field through the double-pass AOM, when the last is driven from the PID fast port at frequency $f_\text{tune}-f_\text{dds}$ (see text); in this case the 2 beams ($900$ MHz apart) are locked relative to each other. b) same as a) but the AOM is driven by an independent fixed DDS. The width is representative of the ECDL linewidth over the time of the 2 s scan. Note the vastly different scales on the horizontal axis.}\label{beat_comparison}
\end{figure}

Next, for better absolute stability, we reference the laser to a highly stable RC without any additional components other than a single AOM (Fig.\ref{full_diagram_0}). A DC voltage in the range of about $0-10$ V is added to the correction signal of the PID to tune the laser to a desired average position relative to the carrier $2f_\text{tune}$. The light reflected from the FC contains non-diffracted light at the frequency of the ECDL carrying sidebands at $30$ MHz (see panel 2 of Fig.\ref{full_diagram_0}). Note that this light still has the noise of the ECDL. It is passed through a double-pass up-diffracting AOM (Crystal Technology, model 3200, 200 MHz center frequency). The AOM is driven by a portion of the radio-frequency signal that comes from the VCO and that is mixed with the signal at $f_\text{dds}$ from a DDS, resulting in an approximately $200$-MHz drive within the range of the AOM ($f_\text{dds}$ is chosen accordingly). The AOM is not a part of the closed loop but since it is driven by the radio-frequency that contains the frequency excursions of the ECDL relative to the FC, it corrects the frequency excursions within its bandwidth. The light after the AOM is shifted by $2f_\text{dds}$ with respect to the light transmitted through the FC (see panel 3 of Fig.\ref{full_diagram_0}). In comparison to the FC-transmitted beam it has enhanced noise beyond the AOM's bandwidth (which is typically $250$ kHz). This, however, is irrelevant to the lock to the RC as the only noise that needs to be compensated for comes from the acoustic noise of the FC.

In Fig.\ref{beat_comparison} we analyze the light that comes out of the AOM. We show the beat signal between the light locked to and transmitted through the FC and the light diffracted by the AOM when the AOM is either driven by an independent source or driven by the signal that is correlated with the signal that is fed to the EOM. When the AOM is driven from a fixed $200$-MHz source the beat as shown in Fig.\ref{beat_comparison} b) reveals the linewidth of the free running ECDL (over the time of the sweep of $2$ s). When driven by the half frequency of the EOM at $f_\text{tune}-f_\text{dds} \sim 200$ MHz the beat signal dramatically narrows as shown in Fig.\ref{beat_comparison} a). We attribute the spurs at multiples of 50 Hz visible in Fig.\ref{beat_comparison} a) to the power supply that drives the LD current source and the PZT drivers (see Fig. \ref{full_diagram_electronics} c)). With appropriate filtering, they could be removed.

\begin{figure}
\includegraphics{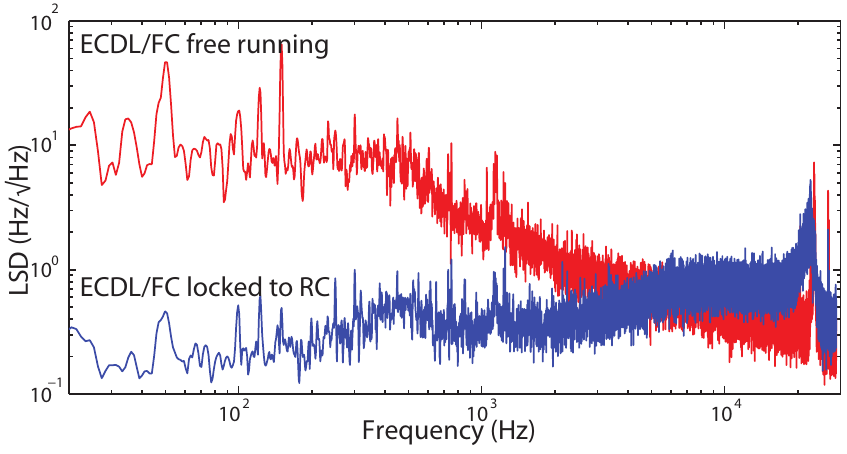}
\caption{Frequency LSD of the PDH error signal as obtained in reflection from the RC when the ECDL/FC system is either tuned to the peak transmission of the RC without the lock via PZTt (free running) or locked to the RC by a single channel PID to the PZTt (locked to RC). }\label{mamacavity}
\end{figure}

With the feedback to the AOM the diffracted light can now be sent to the RC. A single-channel feedback to the fast PZT of the FC (PZTt in Fig.\ref{full_diagram_0}) eliminates the acoustic noise of the FC and provides long-term stability. The light after the AOM already carries PDH-sidebands, eliminating the need for an additional EOM. In Fig.\ref{mamacavity} we compare the LSD of the PDH-signal of the free-running ECDL/FC-system to that of the ECDL/FC-system locked to the RC (cylindrical ULE cavity in vacuum at $10^{-8}$ Torr with finesse $\sim 50000 $). Even without the lock to the RC the free running ECDL/FC laser linewidth as calculated from the LSD is already $\sim 300$ Hz and, as expected, the LSD is significant only in the acoustic range. When locked the acoustic noise is greatly reduced. The laser linewidth relative to the RC is now sub-Hz, although this estimate is based purely on the LSD of the PDH error signal and does not include a possible residual amplitude modulation (RAM) at the PDH frequency. Choosing a smaller (i.e. lighter) cavity mirror and a FC PZT with a smaller inner diameter we expect to be able to increase the BW of the second lock to at least $\sim 50$ kHz and not be limited by the mechanical resonance of the FC PZT-mirror \cite{Briles2010} that can still be seen in the error-signal spectrum.

\section{Conclusions}
\label{sec:4}	
In conclusion, we have demonstrated a simple, compact, and versatile laser system that is easy to operate and that achieves simultaneously high spectral purity, high tunability, long-term stability, and robustness. It should facilitate experiments in high precision spectroscopy, meteorology, interferometry, and quantum state control. We note that this locking scheme does not involve electronic feedback to the laser diode itself. Laser diodes and fibered EOMs cover a large portion of the visible and near-infrared spectrum of interest for many experiments in atomic and molecular physics, making the setup widely applicable. The fibered EOM's highly defined interaction region within the crystal makes the system less susceptible to problems arising from crystal inhomogeneity or anisotropy of the applied field \cite{Kessler2012}. One can also use the EOM for canceling RAM by simply feeding a DC offset on top of the rf correction signal \cite{Wong1985}. An ultimate test for our system would be a beat measurement involving two identical systems locked to independent ultra-stable reference cavities. Our system will first be applied to molecular ground-state transfer on KCs molecules similar to work presented in Ref.'s \cite{Danzl2008,Ni2008,Mark2009,Danzl2010a,Aikawa2011a,Takekoshi2014}.

\section{\label{sec:8}Acknowledgments}
We thank J. Berquist for providing the ULE cavity, J. Danzl for helping with the design of the PID circuit, and K. Aikawa for useful suggestions on the manuscript. We acknowledge generous support by R. Grimm. The work is supported by the European Research Council (ERC) under Project No. 278417.


\begin{thebibliography}{10}
\bibitem{Toptica}
{A commercial ECDL design based on a flexture design is available by the company Toptica (patents DE102007028499 and US7970024).}
\end{thebibliography}
\end{document}